*Dia tōn grammōn*: Euclidean geometry to solve trigonometric problems

Enrico Landi, Francesca Schironi, University of Michigan
Submitted to the Journal of the history of Astronomy on August 18 2023.

# Introduction

Hipparchus (middle of the second century BCE) is the most famous astronomer of the Hellenistic age, but little of his work reached us directly, because the authority of Ptolemy and his *Almagest* led to the loss of the original works written by previous astronomers. The only text written by Hipparchus which survived to the present day by indirect tradition is a polemical commentary aimed at Aratus and his scientific source, Eudoxus, entitled *Exegesis of the Phaenomena of Eudoxus and Aratus*.[1] This treatise is certainly not among Hipparchus' most important scientific works, but it was saved because it was transmitted along with the corpus of Aratus' exegesis, which, on the contrary, was preserved thanks to the great popularity of Aratus' poem *Phaenomena* (ca. 270 BCE) throughout antiquity and the byzantine period.

Despite the title, Hipparchus' *Exegesis* is a complex text with multiple goals. In fact, it does not aim at explaining the poem of Aratus, as the title would suggest. Rather, the first part (1.1.1-2.3.38) of the *Exegesis* is a sharply polemical commentary in which Hipparchus selects all the passages of the *Phaenomena* in which there are mistakes in order to criticize and correct them. The second part (2.4.1-3.4.12), instead, is a self-standing catalogue of simultaneous risings and settings in which Hipparchus provides quite accurate data concerning the rising and setting of the 42 constellations: 16 northern constellations, 14 southern constellations and 12 zodiacal constellations. A third part (3.5.1-3.5.23) identifies a number of stars marking 24 meridians, spaced 1-hour from each other, to help in the calculation of time.

In the catalogue of simultaneous risings and settings each constellation is discussed in two entries, one for the rising and one for the setting, where a number of astronomical data are given; each entry provides the same type of data, described in a precise order. An example of an entry is the following one, about the rising of Perseus (modern star identifications are given in brackets):

> **2.5.15** Τοῦ δὲ Περσέως ἀνατέλλοντος συνανατέλλει μὲν αὐτῷ ὁ ζῳδιακὸς ἀπὸ Αἰγόκερω ε$^{ης}$ καὶ κ$^{ης}$ μοίρας ἕως Κριοῦ ιδ$^{ης}$ μέσης· μεσουρανεῖ δὲ ἀπὸ Σκορπίου ι$\varsigma^{ης}$ μέσης ἕως Αἰγόκερω η$^{ης}$ μέσης. καὶ πρῶτος μὲν ἀστὴρ ἀνατέλλει ὁ ἐν τῇ ἅρπῃ νεφελοειδής, ἔσχατοι δὲ οἱ ὑπὲρ τὴν Πλειάδα ἐν τῷ ἀριστερῷ ποδὶ κείμενοι.
> Μεσουρανεῖ δὲ ἀστὴρ πρῶτος μὲν ὁ ἐν μέσῳ τῷ Θυμιατηρίῳ λαμπρός, καὶ τοῦ Ἐνγόνασιν ὁ προηγούμενος τοῦ δεξιοῦ ὤμου ἐν τῷ βραχίονι, ἔσχατος δὲ τοῦ Αἰγόκερω ὁ βορειότερος τῶν ἐν τοῖς γονατίοις, καὶ τοῦ Ὄρνιθος ὁ βορειότερος τῶν ἐν τῇ δεξιᾷ πτέρυγι, ὡς ἡμιπήχιον προηγούμενος τοῦ μεσημβρινοῦ.
Ἀνατέλλει δὲ ὁ Περσεὺς ἐν ὥραις τρισὶ καὶ ἡμίσει καὶ τρίτῳ μέρει ὥρας.

---

[1] The only modern edition (with a German translation) is Manitius 1894; Schironi is preparing an edition with translation and commentary (Schironi (forthcoming) ). The passages from the Exegesis presented below, as well as the translation, are taken from this forthcoming work.

At the rising of Perseus, the zodiacal [circle] rises together with him from the 25th degree of Capricorn to the middle of the 14th [degree] of the Ram. In mid-sky is [the section of the zodiacal circle] from the middle of the 16th [degree] of the Scorpion to the middle of the 8th [degree] of Capricorn. And the first star to rise is the cloud-like [star] in the sickle (NGC 869 + 884); the last [to rise] are the [stars] which lie in his left foot, over the Pleiad (ζ Per, o Per).
The first star to be in mid-sky is: the bright [star] in the middle of the Incense-Burner (α Ara); and, of the Kneeler, the [star] that, in the arm, precedes the right shoulder (γ Her). The last [to be in mid-sky] is: of Capricorn, the more northern [star] of those in the small knees (ψ Cap); and, of the Bird, the more northern of the [stars] in the right wing (κ Cyg), which precedes the meridian by ca. half a cubit.
Perseus rises in three hours and 5/6 of an hour [lit.: half of an hour and a third of an hour].

The data provided by this (and every other) entry can be grouped in three categories as follows, distinguishing between rising (or setting) data, culmination data, and event duration:

First set of data (rising/setting):

1. Arc of the ecliptic rising (or setting) together with the rising (or setting) constellation;
2. The first and last stars of the constellation to rise (or set).

Second set of data (culmination):
1. Arc of the ecliptic culminating together with the rising (or setting) constellation;
2. A series of stars that culminate when the first and last stars of the constellation rise (or set).

Third set of data (duration):
1. The time it takes for the constellation to rise (or set), in equinoctial hours.

We know Hipparchus had a Catalogue of Stars in equatorial coordinates,[2] which he also uses through the *Exegesis* to identify the position of stars.[3] Also, Hipparchus most likely used a globe to determine which stars were the first and last star to rise (or set) for each constellation, as well as to identify which stars were culminating when a specific star was either rising or setting.
The question is how he could determine the ecliptic data. The most natural answer is that he used the globe to determine those data as well, by providing it with a well calibrated, graduated ecliptic circle. This was the quickest and easiest option, and most likely the one he used to compile his catalogue of simultaneous risings and settings. However, Hipparchus explicitly states (2.2.24-27) that in a *Treatise on the Simultaneous Risings* (ἡ τῶν συνανατολῶν πραγματεία) he developed a method to determine which point on the ecliptic is rising/setting and culminating simultaneously with a star which is rising or setting. This treatise is lost to us, and in the *Exegesis* Hipparchus does

---

[2] See Gysembergh, Williams and Zingg 2022.
[3] See Duke 2002; Schironi (forthcoming), Chapter 5.

not provide the details of his proof; still, he gives one short example: the setting of υ Boo. In particular, he outlines the following steps:

**2.2.27** ἀλλ' ὁ εἰρημένος ἀστὴρ κεῖται ὡς κατὰ παράλληλον κύκλον περὶ <τὴν> α$^{ην}$ μοῖραν τῶν Χηλῶν. δύνοντος ἄρα αὐτοῦ δεῖ μεσουρανεῖν ὡς κατὰ παράλληλον κύκλον τὴν κα$^{ην}$ δ′ [sic; leg. τὴν κδ$^{ην}$ vel κ$^{ην}$ καὶ δ$^{ην}$] τοῦ Αἰγόκερω· ταύτης δὲ ἐπὶ τοῦ παραλλήλου μεσουρανούσης δεῖ ἐν τῷ ζῳδιακῷ μεσουρανεῖν τὴν β$^{αν}$ καὶ κ$^{ην}$ μοῖραν τοῦ Αἰγόκερω. ἀλλ' ἐν τῷ ζῳδιακῷ τῆς β$^{ας}$ καὶ κ$^{ης}$ τοῦ Αἰγόκερω μοίρας μεσουρανούσης, ἐν τῷ ὑποκειμένῳ ὁρίζοντι δεῖ ἀνατέλλειν τὴν ϛ$^{ην}$ μάλιστα μοῖραν τοῦ Ταύρου

But the star in question lies on the parallel circle around <the> 1st degree of the Claws. When it sets, then the 21st and ¼ degree [sic; leg. the 24th degree][4] of Capricorn on the parallel circle must be in mid-sky. But when this [degree of Capricorn] is in mid-sky on the parallel [circle], then the 22nd degree of Capricorn in the zodiacal [circle] must be in mid-sky. But when the 22nd degree of Capricorn in the zodiacal [circle] is in mid-sky, in the horizon that we assume [i.e., Rhodes] roughly the 6th degree of the Bull must rise.

In other words, Hipparchus calculates three different sets of data, reporting them in the following order:

1. Determining what degree of the celestial equator is culminating at the time of the star's setting (once the star's declination and right ascension are known).
2. Calculating the corresponding degree of the ecliptic which culminates at the same time.
3. Calculating the degree of the ecliptic rising at the same time.

Each step uses the results of the previous one. Since Hipparchus does not say how he was able to calculate these parameters, scholars are left to speculations.

## Methodological layout

Hipparchus' method can be broken down following the description he gives in the commentary, identifying three subsequent problems:

A. To calculate the angular length of the portion of the star's parallel located above the horizon;
B. To calculate the culminating point of the ecliptic;
C. To calculate the rising point of the ecliptic.

This problem and Hipparchus' solution to it have already been studied by Sidoli in 2004.[5] He considers (B) to be the primary problem, and (A) and (C) as secondary problems; he addresses (A) using the analemma construction, and solves both (B) and (C) by applying the Menelaus Theorem to two different arc configurations. Mathematically, Sidoli's reconstruction is correct, and has the advantage to be relatively easy to implement and involve only a few steps of calculation, even though he does not describe in detail how the analemma solution to problem (A) was calculated. However, Sidoli's solution to problems (B) and (C) heavily relies on the availability of Menelaus' Theorem at the time of Hipparchus, for which we do not have any direct evidence, since it is mentioned by Ptolemy.[6] In fact, if the theorem originated with Menelaus, it would be later than Hipparchus since the former is dated to the first century CE (Ptolemy reports an observation by him in 98 CE). On the other hand, Toomer[7] briefly addressed this problem without using the Menelaus Theorem; rather, he suggested the use of the analemma to solve problem A, and then added that the solution of the simultaneous risings and settings (problems B and C) was likely achieved using stereographic projections, although no explanations are given on how in practice he could reach certain results. Also, Hipparchus' knowledge of stereographic projections is debated.[8]

---

[4] Corrections of numbers are not directly inserted in the text in Schironi's text when the manuscript tradition unanimously transmits one reading; the reading is kept and in parenthesis the astronomically correct number is given.
[5] Sidoli 2004.

In this paper we thus try to address the following question: could such a problem be solved using less advanced mathematical tools, that is, dispensing from the use of Menelaus' Theorem or stereographic projections? In the following discussion, we show that such a solution is possible utilizing only three tools:

A. A Table of Chords (and the ability of interpolating it)
B. The Theory of Proportions
C. The Pythagoras' Theorem

Both (B) and (C) were certainly known at the time of Hipparchus, while we know that he developed his own Table of Chords (henceforth TOC), apparently with an angular resolution of 7.5 degrees.[9] Such a table could easily (albeit a little tediously) be calculated by means of applying the Pythagoras' Theorem to a circle of known radius R to determine the following relationships:

---

[6] See Neugebauer 1975, 26-29; Pedersen and Jones 2011, 72-78. On ancient spherical trigonometry and Menelaus' Theorem, see also van Brummelen 2009; van Brummelen 2020.

[7] Toomer 1978, 210-211

[8] The evidence is the following passage by Synesius of Cyrene (ca. 370-413 CE),: *On the Astrolabe* 5.1-3 (= Hipp. Fr. 63 Dicks): Σφαιρικῆς ἐπιφανείας ἐξάπλωσιν, ταυτότητα λόγων ἐν ἑτερότητι τῶν σχημάτων τηροῦσαν, ἠνίξατο μὲν Ἵππαρχος ὁ παμπάλαιος, καὶ ἐπέθετό γε πρῶτος τῷ σκέμματι· ἡμεῖς δέ, εἰ μὴ μεῖζον ἢ καθ' ἡμᾶς εἰπεῖν, ἐξυφήναμέν τε ἄχρι τῶν κρασπέδων αὐτὸ καὶ ἐτελειώσαμεν, ἐν πλείστῳ δή τινι τῷ μεταξὺ χρόνῳ τοῦ προβλήματος ἀμεληθέντος, Πτολεμαίου τοῦ πάνυ καὶ τοῦ θεσπεσίου θιάσου τῶν διαδεξαμένων αὐτὴν μόνην ἔχειν ἀγαπησάντων τὴν χρείαν, ἣν ἀρκοῦσαν εἰς τὸ νυκτερινὸν ὡροσκοπεῖον οἱ ἑκκαίδεκα ἀστέρες παρείχοντο, οὓς μόνους Ἵππαρχος μετατιθεὶς ἐγκατέταξε τῷ ὀργάνῳ. [Hipparchus, who lived a long time ago, expressed himself obscurely about the unfolding of a spherical surface that maintains the identity of the proportions in a different form, and was the first to apply himself to such problem. But we, if it is not too much for us to say, have finished weaving it up to the borders and have perfected it, since the problem had been neglected in most of the intervening time; the great Ptolemy and the divine company of his successors were content just to use it, as the 16 stars, which were the only ones that Hipparchus transposed and placed in the instrument, allowed [this instrument] to be enough for a nightly clock]. Neugebauer 1975, 301-304 and 868-869 believed that Hipparchus was well acquainted with the methods of the analemma and stereographic projections, and even that he might be their inventor. However the testimony of Synesius is not very clear about the methodology developed by Hipparchus and is not supported by any other source; for example, Ptolemy never speaks of stereographic projection or plane astrolabes when talking of Hipparchus; cf. Dicks 1960, 204-207.

[9] Cf. Toomer 1974, 6.

$$crd(\pi - \alpha) = \sqrt{4R^2 - crd^2(\alpha)}$$
$$crd\left(\frac{\alpha}{2}\right) = \sqrt{2R^2 - R \times crd(\pi - \alpha)}$$

Enrico Landi has calculated, pen and pencil, such a table using Heron's method and the digit-by-digit calculation in less than an afternoon.

It is extremely important to recall that the TOC was a table of segments, not of a dimensionless trigonometric function such as modern cosine, sine and the like, and as such it required the definition of the radius R. For example, Ptolemy assumes R = 60 (from a diameter divided in "in 120 parts"; see *Almagest* I 9, vol. 1.1, 31.16 Heiberg), a rather convenient choice when using a sexagesimal system. Hipparchus, on the contrary, seems to have adopted the less obvious value R = 3438', a value that seems to have passed into Indian astronomy as well[10] and which we will not discuss here.

The method we will discuss below was certainly much slower and significantly more computationally intensive than the one proposed by Sidoli, and there is little doubt that once Menelaus' Theorem was introduced, astronomers quickly switched to using the latter. It is even possible that some of its most cumbersome calculations could have been simplified using theorems that we do not consider in this paper, in our attempt to utilize the simplest geometrical tools available. Still, the following method is geometrically exact and makes use of tools unquestionably available at the time of Hipparchus. Most importantly, it reduces a problem of calculating the relationships of angles on the sphere to a series of problems of plane geometry, sparing Hellenistic astronomers the uncomfortable need to develop a new mathematical theory (equivalent to modern trigonometry), while at the same time allowing them to use the branch of mathematics they were most comfortable with and which they had already developed to a considerable level of sophistication: plane geometry.

## Input data

One of the most attractive features of the methodology developed by Hipparchus was that, as he himself states, it could be applied to calculate rising, culminations and settings almost anywhere in the world:

> **2.4.3** τὰς δὲ κατὰ μέρος αὐτῶν ἀποδείξεις ἐν ἄλλοις συντετάχαμεν οὕτως, ὥστε ἐν παντὶ τόπῳ σχεδὸν τῆς οἰκουμένης δύνασθαι παρακολουθεῖν ταῖς διαφοραῖς τῶν συνανατολῶν καὶ συγκαταδύσεων.
>
> We have put together detailed demonstrations for these [questions] in other [works], so that in almost every place of the inhabited world it is possible to keep up with the differences in the simultaneous risings and settings.

In other words, this methodology could be used by an observer located at any latitude and provide the risings, settings and culmination observed locally. We can speculate what Hipparchus meant by "almost anywhere in the world" since this methodology applies anywhere you can observe a star rising and setting, and the only place where it fails is where there are no risings and settings at all — namely, the equatorial north and south pole.

---

[10] See Neugebauer 1972, 249-251; Neugebauer 1975, 299-300, and Toomer 1974.

In any case, this methodology requires the user to know three sets of data before starting the calculation:

1. The star's right ascension and declination
2. The location's latitude in the equatorial reference system
3. The *inclination of the cosmos*, namely the obliquity of the ecliptic.

We already know that Hipparchus had his own stellar catalogue, for which he must have had a set of equatorial coordinates; it is beyond the scope of this paper to discuss such a catalogue.[11] He also had means available to calculate the latitude of the place he was referring his calculations to, namely, Rhodes.[12] Furthermore, he also had means to calculate the obliquity of the ecliptic; Ptolemy (*Almagest*, 1.12, vol. 1.1.67.17-68.6 Heiberg) states that Hipparchus used the same value as found by Eratosthenes and which was close to the value which Ptolemy himself derived: half of 11/83 of the circumference (corresponding to 23º 51', against the value of the time of 23º 42', obtained from the reconstructions of the programs Skyfield and Stellarium). In the *Exegesis* however Hipparchus himself states he adopted a value of 24º (**1.10.2** μὲν γὰρ θερινὸς τροπικὸς τοῦ ἰσημερινοῦ βορειότερός ἐστι μοίραις ὡς ἔγγιστα κδ̄ [for the summer tropic is farther to the north than the equator by approximately 24 degrees]).

## First problem — Calculate the length of the portion of the star's parallel located above the horizon

This problem is best solved through the use of the analemma, so we solve this step in the same way as Toomer and Sidoli. We will describe here how the analemma could be tackled using only the three mathematical tools we utilize in this discussion. In particular, the analemma can be easily solved using only a TOC and the Theory of Proportions, once the latitude ψ of the observer's location is known, without even (directly) using the Pythagoras' theorem, nor the obliquity of the ecliptic.

Figure 1 displays a 2D rendering of the geometry of the analemma, for a star in the northern hemisphere (positive declination δ, top) and in the southern hemisphere (negative declination δ, bottom); in both cases the methodology to follow is the same. Figure 1 shows the celestial sphere as seen from the west direction: the plane of the celestial equator and of its parallels are shown as straight lines and the south meridian is shown by the circle ABDCM centered in O. The Earth rotation axis is inclined relative to the local horizon by an angle ψ, the local latitude. The parallel with declination δ where the star in question orbits is shown by the segment BD, the star sets (and rises) in point G and culminates in point D; point F is the intersection between the parallel plane where the star is and the Earth's axis. The red curve represents the 90º projection of the parallel BD into the plane of the south meridian: point L corresponds to point F, and point H corresponds to point G and represents the position of the star on the parallel at the time of setting (or rising). Point E is the projection of point D onto the diameter AC of the celestial sphere.

---

[11] See Gysembergh, Williams and Zingg 2022.

[12] As he clearly states at the beginning of its catalogue of simultaneous risings and settings in the Exegesis: **2.4.2** ἑξῆς δὲ ὑποτάξω περὶ ἑκάστου τῶν ἀπλανῶν ἄστρων ἐπὶ κεφαλαίου, τίνι τε τῶν δώδεκα ζῳδίων συνανατέλλει καὶ συγκαταδύνει, καὶ ἀπὸ πόστης μοίρας τοῦ ζῳδίου ἀρξάμενον ἕως πόστης μοίρας ἔσχατον ἀνατέλλει ἢ {συγκατα}δύνει ἐν τοῖς περὶ τὴν Ἑλλάδα τόποις καὶ **καθόλου ὅπου ἐστὶν ἡ μεγίστη ἡμέρα ὡρῶν ἰσημερινῶν ῑδ̄ καὶ ἡμιωρίου** [Next, for each of the fixed constellations, I will outline, in short, 1) together with which of the twelve zodiacal signs it rises and sets, and 2) from which degree of the zodiacal sign it starts and 3) at which degree it ends its rising and setting in the places around Greece, and, **in general, where the longest day consists of 14 and a half equinoctial hours [ = Rhodes]]**. That this place is Rhodes is clarified by the following passage: **1.7.21** ὅπου δέ ἐστιν ἡ μεγίστη ἡμέρα ὡρῶν ῑδ̄ ∠, ἐκεῖ ὁ ἀεὶ φανερὸς κύκλος ἀπέχει ἀπὸ τοῦ πόλου μοίρας λ̄ς̄, ἐν Ἀθήναις δὲ μοίρας λ̄ζ̄ [**Where the longest day is 14 and ½ hours, there the ever-visible circle is 36 degrees away from the pole** and in Athens [it is] 37 degrees [away from the pole]. ]

The portion of the star's parallel located above the horizon is twice the circular section HLD, and corresponds to an angle twice as large as the angle HFD. The angle HFD will be obtained by adding or subtracting the angle α to 90 degrees in the case of positive and negative declination, respectively (see Figure 1):

$$H\hat{F}D = \begin{cases} 90º + \alpha & \text{with} \quad \delta > 0 \\ 90º - \alpha & \text{with} \quad \delta < 0 \end{cases}$$

The calculation of the angle HFD can be carried out in three steps considering Figure 1 and solving for the segment GF using angles δ, ψ (the star declination and the local latitude) first, and then using the TOC to determine the angle α.

First, assuming the radius OD of the circle to have the same value R of the TOC, the segments OF and HF can be readily calculated applying the TOC (or by interpolating from it) to the triangle ODE:

$$DF = HF = \frac{1}{2}crd\left[2\left(\frac{\pi}{2} - \delta\right)\right] \qquad DE = OF = \frac{1}{2}crd(2\delta)$$

Next, by applying the TOC to triangle OFG, we have

$$\frac{OG}{OF} = \frac{R}{\frac{1}{2}crd\left[2\left(\frac{\pi}{2} - \psi\right)\right]} \qquad \frac{GF}{OG} = \frac{\frac{1}{2}crd(2\psi)}{R}$$

from which the segment GF can be readily calculated. This step involves the Theory of Proportions. In fact, it is very important to note that the triangle OFG can be inscribed in a circle with center O and radius OG, which is smaller than the radius R used for the TOC. Thus, in order to apply the TOC to triangle OFG, it is necessary to apply the Theory of Proportions between the triangle OFG and a similar triangle where the equivalent to the segment OG has length R. This step is necessary because of the nature of the TOC itself as a table of *segments*, unlike the modern trigonometric functions, so that the values of these segments depend on the value of the radius of the circle for which the TOC has been calculated. This application of the Theory of Proportion is necessary every time the TOC is applied to a triangle rectangle whose hypothenuse is different from R, and is implicit in all the uses of the TOC we do in this work.

The third and last step yields directly the angle α: knowing HF and applying again the TOC to triangle HGF, we have

$$\frac{GF}{HF} = \frac{\frac{1}{2}crd(2\alpha)}{R}$$

which allows us to determine the value of the angle α from the TOC.
To check the consistency with modern formulas, we can apply modern algebra and manipulate these steps to obtain the formula

$$crd(2\alpha) = 2R \frac{crd(2\delta)}{crd(\pi - 2\delta)} \frac{crd(2\psi)}{crd(\pi - 2\psi)}$$

which is equivalent to the modern formula sinα = tanδ tanψ obtained noting that

$$R \sin \alpha = \frac{1}{2} crd(2\alpha)$$

As noted earlier, once the angle α is known, the length of the parallel above the horizon is given by L = 2 x HFD and the culminating point of the equator $RA_{culm}$ can be readily calculated from the right ascension $RA_{star}$ of the star as

$$RA_{culm} = RA_{star} + H\hat{F}D = RA_{star} + \begin{cases} 90° + \alpha & \text{with} \quad \delta > 0 \\ 90° - \alpha & \text{with} \quad \delta < 0 \end{cases}$$

## Second problem — Calculate the culminating point of the ecliptic

According to Sidoli, this is the primary problem[13] of Hipparchus' *Dia Tōn Grammōn* method, but in reality it boils down to a simple change of coordinates between the Equatorial system and the Ecliptic system, for a point lying on the Equator, only applied to the ecliptic longitude (that is, the ecliptic latitude is of no interest). Sidoli applies the Menelaus Theorem for the first time to solve it,[14] but it is possible to reach the solution by means of the TOC, the Theory of Proportions, and the Pythagoras' Theorem, once the obliquity ε is known.

Finding the ecliptic longitude of point C of the ecliptic sitting on the same meridian of point B on the celestial equator means solving a problem set as shown in Figure 2, where A is the Spring Equinox. The geometry of this problem is the same for both equinoxes, and for RA angles preceding or following the Equinox itself.

Figure 2 shows a 3D rendering of the problem. Both panels show a section of the celestial sphere, centered in O with radius OA = OB = OC = R and A is the spring equinox. The section of the circle OAB lies on the plane of the celestial equator, so that the arc AB is the section of the celestial equator corresponding to the culminating right ascension $RA_{culm}$ calculated in the previous section. The section of the circle OAC lies on the ecliptic plane, so that the arc AC is the section of the ecliptic corresponding to the culminating ecliptic longitude $\lambda_{culm}$; the arc BC belongs to the local south meridian. The segments AC" and AB" are tangent to the ecliptic and the celestial equator, respectively, and are delimited in C" and B" by the extensions of radiuses OC and OB, respectively; the segment B"C" is perpendicular to AB".

---
[13] Cf. Sidoli 2004, 73.
[14] Sidoli 2004, 75-76.

From Step 1, we have calculated the culminating right ascension $RA_{culm}$ corresponding to the setting of a star, so $RA_{culm}$ is the right ascension of point B. The problem consists of determining the corresponding culminating ecliptic longitude $\lambda_{culm}$ as the point of the ecliptic sitting on the culminating meridian. The problem can be solved considering in series three triangles in Figure 2: the *Equatorial triangle OAB"*, the *Connection triangle AB"C"*, and the *Ecliptic triangle OAC"*. Triangles A'BC' and A'''B'''C are parallel to the Connection Triangle, with a vertex in B and C, respectively. We use the same TOC used in the First Problem, applying it to a sphere with the same radius R as the TOC.

We start from the Equatorial triangle (Figure 2, top). Using the TOC and the $RA_{culm}$ angle determined in the first problem, we have

$$A'B = \frac{1}{2}crd(2RA_{culm}) \qquad OA' = \frac{1}{2}crd\left[2\left(\frac{\pi}{2} - RA_{culm}\right)\right]$$

Applying the TOC to triangle OAB", noting that $OA = OB = R$, we can readily calculate $AB''$:

$$\frac{A'B}{OA'} = \frac{AB''}{OA} \quad \longrightarrow \quad AB'' = R\frac{crd(2RA_{culm})}{crd(\pi - 2RA_{culm})}$$

We now move to the Connection triangle, noting that through the application of the TOC to the Connection triangle itself we can determine AC":

$$\frac{AB''}{AC''} = \frac{\frac{1}{2}crd\left[2\left(\frac{\pi}{2} - \varepsilon\right)\right]}{R} \quad \longrightarrow \quad AC'' = AB''\frac{R}{\frac{1}{2}crd(\pi - 2\varepsilon)}$$

As a sanity check, manipulating the result for AC" using modern algebra, we obtain the modern trigonometric formula

$$\tan(\lambda_{culm}) = \frac{\tan(RA_{culm})}{\cos \varepsilon}$$

However, the lack of a conceptual framework where 1) the geometrical relationships within a circle do not depend on the radius of the circle itself (like in trigonometry); and 2) relationships between quantities can be expressed by functions, forced ancient astronomers to devise a method to calculate the actual length of the segment corresponding to the angle $\lambda_{\text{culm}}$, which allowed them to use the TOC and determine the angle $\lambda_{\text{culm}}$ itself.

Knowing AC″ allows us to move into the Ecliptic triangle, where we need to first calculate OC″ through the Pythagoras' theorem:

$$OC'' = \sqrt{R^2 + (AC'')^2}$$

Once OC″ is known, then we can apply the Theory of Proportions to triangles OAC″ and OA‴C to determine the segment A‴C, to yield the relationship

$$\frac{AC''}{OC''} = \frac{A'''C}{R} \quad \longrightarrow \quad A'''C = \frac{1}{2}crd(2\lambda_{culm}) = R\frac{AC''}{OC''}$$

from which the TOC allows the determination of angle $\lambda_{culm}$.

It is important to note that this geometrical setup also allows the user to walk the opposite route: knowing the angle $\lambda_{culm}$, the $RA_{culm}$ can be determined.

### Third problem — Calculate the rising point of the ecliptic

Once the culminating point of the ecliptic $\lambda_{culm}$ is determined, we are left with the last step of Hipparchus' method, which yields the rising point of the ecliptic. Since the ecliptic is a great circle, this method can actually provide both the rising and setting point of the ecliptic for a given value of $\lambda_{culm}$, since they will be simply separated by 180 degrees. This is where Menelaus Theorem, again invoked by Sidoli as the solution,[15] is most useful: in fact, to solve this problem using the simpler geometrical tools as we do in this paper involves two steps, each requiring a long series of calculations. The first step consists of determining the arc of the horizon between the rising (or setting) point of the ecliptic and the local south direction; in the second step the length of the ecliptic from rising to culmination is determined through a change of coordinates from the horizontal to the ecliptic system, using the same method as in Step 2 but considering that the inclination ε' of the ecliptic axis relative to the local vertical changes during the day due to the Earth's own rotation. While the Menelaus Theorem as applied by Sidoli allows for the solution in one simple, neat step, with the following geometrical method both steps are necessary each involving multiple calculations.

     Sidoli considered this third step as a secondary problem, but in reality it is the most geometrically complex and computationally expensive, making its application rather cumbersome. Still, despite the complexity, this problem can again be solved through the use of the TOC, the Theory of Proportions, and the Pythagoras' theorem.

     To solve this problem it is useful to refer to Figure 3, which shows 3D renderings of the celestial sphere centered in O as seen from four different vantage points: from the local south (top left), the Zenith (top right), the local west direction (bottom left), and from an unspecified direction in the southeast quadrant. In all panels, the circle ESWN is the local horizon, point Z is

---

[15] Sidoli 2004, 80-81.

the local Zenit, and the SZN circle is the local south meridian. The blue curve in the bottom right panel is the half of the ecliptic above the horizon, culminating in point C. The ecliptic — a great circle — intersects the horizon in the two points E' and W' (shown in all panels) separated by 180 degrees; points N' and S' are the intersections of the diameter of the plane of the horizon perpendicular to the E'W' segment, and θ is the angle between the N'S' diameter with the N-S direction. The circle centered in O' (dotted in the top left panel, full line in the top right panel, seen as a segment in the bottom left panel) is the circle made by the ecliptic N pole on the celestial sphere around the Earth's rotation axis (where the O'O segment lies); point A represents the position of the ecliptic N pole in this circle and is indicated by the angle φ (see below for its relationship to the RA), which we assume to be zero when the summer solstice culminates (corresponding to RA = 6h00m, λ = 90°) and changes at constant speed following the Earth's own rotation, covering 360 degrees in 24 hours. Thus, the angle φ is the same as the RA, but shifted by 6h, or 90 degrees.

**First step.** In order to determine the points of the ecliptic that rise and set at any given time it is necessary to determine the position S', whose distance from the local south S is indicated by the angle θ. The length of the arc spanning from the position E' of the ecliptic rising point (or from the ecliptic setting point, W') to the local south direction S can then be easily determined by adding the angle θ to, or subtracting it from, 90 degrees (depending on the time of the day).

To calculate the angle θ it is first important to note that there are a few fixed segments involved in the procedure, that can be calculated once and for all and applied the rising or setting of any star. These are the radius O'A of the circle made by the rotation of the ecliptic N pole, the distance O'O between the center of that circle to the center O of the celestial sphere, and the projection O"O of the O'O segment on the local vertical direction. Using a TOC based on the radius R of the celestial sphere:

$$\begin{cases} O'A = \frac{1}{2}crd(2\varepsilon) \\ O'O = \frac{1}{2}crd\left[2\left(\frac{\pi}{2} - \varepsilon\right)\right] \\ O''O = \frac{O'O}{R}\frac{1}{2}crd(2\psi) \end{cases}$$

This step can be solved considering the segment AB inside the circle of the ecliptic N pole. In that circle, the segment AB can be expressed through the chords of both angles φ and θ: giving us the means to determine the latter through the former:

$$\begin{cases} AB = O'A\,\dfrac{\frac{1}{2}crd(2\varphi)}{R} \\ AB = O'''A\,\dfrac{\frac{1}{2}crd(2\theta)}{R} \end{cases}$$

In both cases, we have applied the TOC to the triangles OO'A (equation involving φ) and OAO''' (equation involving θ).

The segment AB can be readily determined noting that O'A is known, and that the angle φ is related to the culminating right ascension $RA_{culm}$ by

$$\varphi = RA_{culm} - 90^0$$

O'''A is determined by subtracting (see Figure 3, bottom left) the segment O''O''' = O'R from the known segment O''O to obtain OO''', and then using the Pythagoras' theorem on the triangle OAO'''. This can be done in the following steps.

First, noting that the angle RO'B is the complementary to the latitude, the application of the TOC on the triangle AO'B yields:

$$O'B = O'A \frac{\frac{1}{2}crd\left[2\left(\frac{\pi}{2} - \varphi\right)\right]}{R}$$

Once O'B is known, in the triangle RO'B we can use the TOC calculate O'R as:

$$O'R = O'B \frac{\frac{1}{2}crd\left[2\left(\frac{\pi}{2} - \psi\right)\right]}{R}$$

from which

$$OO''' = OO'' - O'R$$
$$O'''A = \sqrt{R^2 - (OO''')^2}$$

At this point, the angle θ can be determined from the TOC and using the segments O'A and O'''A just obtained, and from this we can determine the length of the arc E'S (in degrees) which is the projection of the arc of the ecliptic E'C spanning from rising to culmination, which is the last piece of information we need to determine the rising point of the ecliptic.

**Second step.** This step is needed to determine E'C, for which it is convenient to refer to Figure 3 (bottom right). This figure shows that the configuration between E'C and E'S is the same as the configuration shown in Figure 2 when we changed the coordinate system from equatorial to ecliptic, with two main differences:

A. The equatorial system is now replaced by the horizontal system; and
B. The inclination of the ecliptic over the horizontal plane, ε', depends on time and is different from the value of ε, except when the equinoxes culminate

However, in Figure 3 the direction of the ecliptic axis is indicated by the segment OA, so that the new inclination of the ecliptic can be determined applying the TOC to the triangle OO'''A, noting that

$$O'''A = \frac{1}{2}crd(2\varepsilon')$$

Thus, once ε' is determined, we can repeat the procedure of Step 2 to derive the arc E'C (length of ecliptic from rising to culmination) from the arc E'S (length of the horizon from ecliptic rising to southern meridian). The result can be subtracted from $\lambda_{culm}$ to obtain $\lambda_{rise}$. Since both the ecliptic and the horizon are great circles, the setting ecliptic point $\lambda_{set}$ can be determined by simply adding 180 degrees to $\lambda_{rise}$.

It is important to notice that the method of Step 3 can be equally used to determine $\lambda_{set}$ first in place of $\lambda_{rise}$, but needs to be applied to the calculation of the arc between E'S and W'S which is smaller than 90 degrees, so that the geometry of the change of coordinates described in Step 2 can be applied.

## Conclusions

Although the method we described to calculate the simultaneous risings and settings of a star is mathematically exact, it is by no means to be understood as the method that Hipparchus himself was referring to. There is no evidence that this method is indeed Hipparchus' own method, especially because the treatise he wrote on the subject has been lost. All we wanted to stress is that there is no need of the Menelaus theorem or stereographic projections to solve this problem, and that more basic computational tools, certainly at Hipparchus' disposal, can equally lead to the answer, even though through a much more computationally expensive procedure. It is very well possible that the present solution to this problem could be simplified by the use of geometrical theorems available to Hipparchus, which either we did not consider or for which we do not know the existence: our goal was to show that all the steps in the problem (analemma, change of coordinates to and from a spherical system and another, and even the rotation of the sky) could be successfully addressed with very basic geometrical tools unquestionably available to Hipparchus.

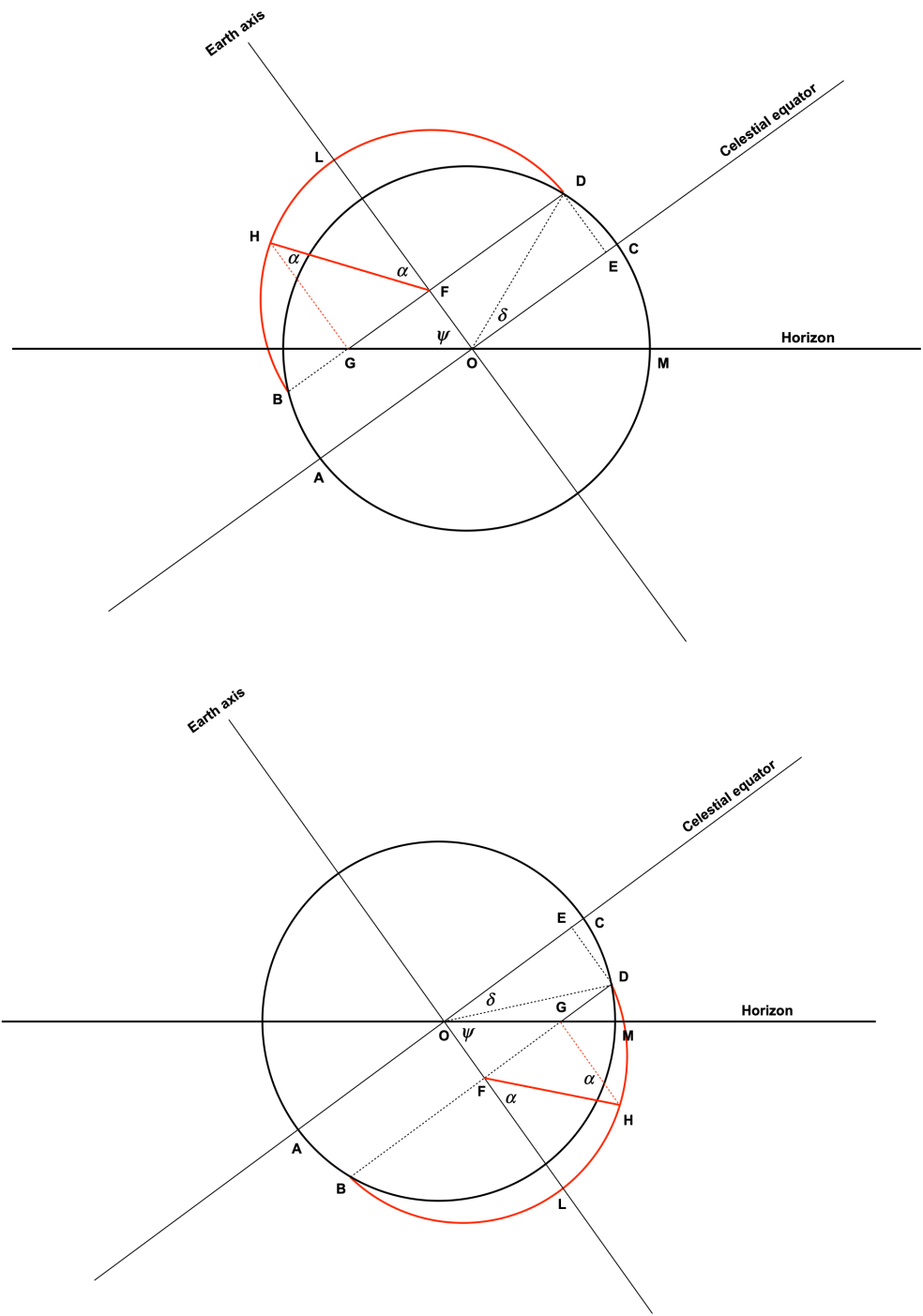

**Figure 1**. Analemma configuration for a star in the Northern (top) and Southern (bottom) hemisphere. The black circle (of radius OC) is the section of the celestial sphere as seen from the West direction; the Horizon corresponds to a local latitude ψ. The star (having a declination δ and setting at point G), rotates around the Earth's axis along the parallel BD (seen as a segment); the red curve BLD represents the projection of the parallel BD on the plane of the black circle, and the point H represents the position of the star in the projected parallel. Point E is the projection of the parallel's culmination point D on the equatorial plane.

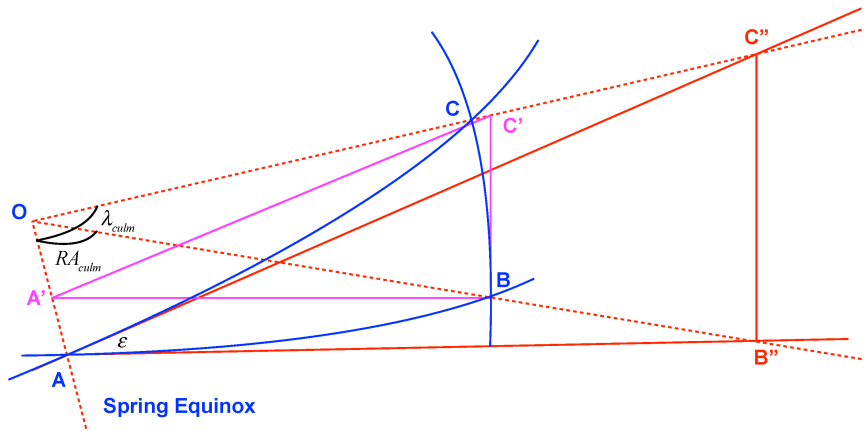

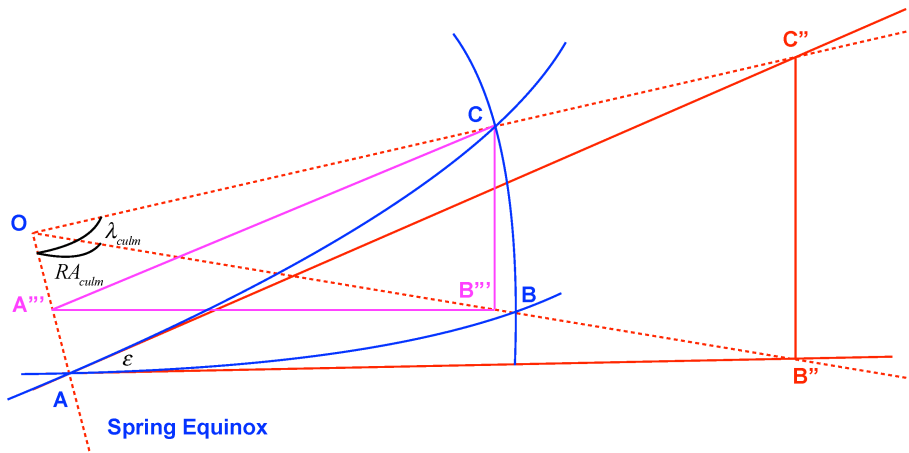

**Figure 2.** Change of coordinates close to the Spring Equinox (A). $RA_{culm}$: Culminating right ascension angle; $\lambda_{culm}$ : culminating ecliptic longitude; $\varepsilon$: obliquity of the ecliptic. AB is the arc of the celestial equator corresponding to the angle $RA_{culm}$ ; AC is the corresponding arc on the ecliptic (subtending the angle $\lambda_{culm}$); CB is the South meridian. The Connection triangle is given by AB"C". Top: working on the Equatorial triangle $OAB"$ , where A'BC' is a triangle similar to the Connection Triangle touching point B; Bottom, working on the Ecliptic triangle $OAC"$ i triangle similar to the Connection Triangle touching point C.

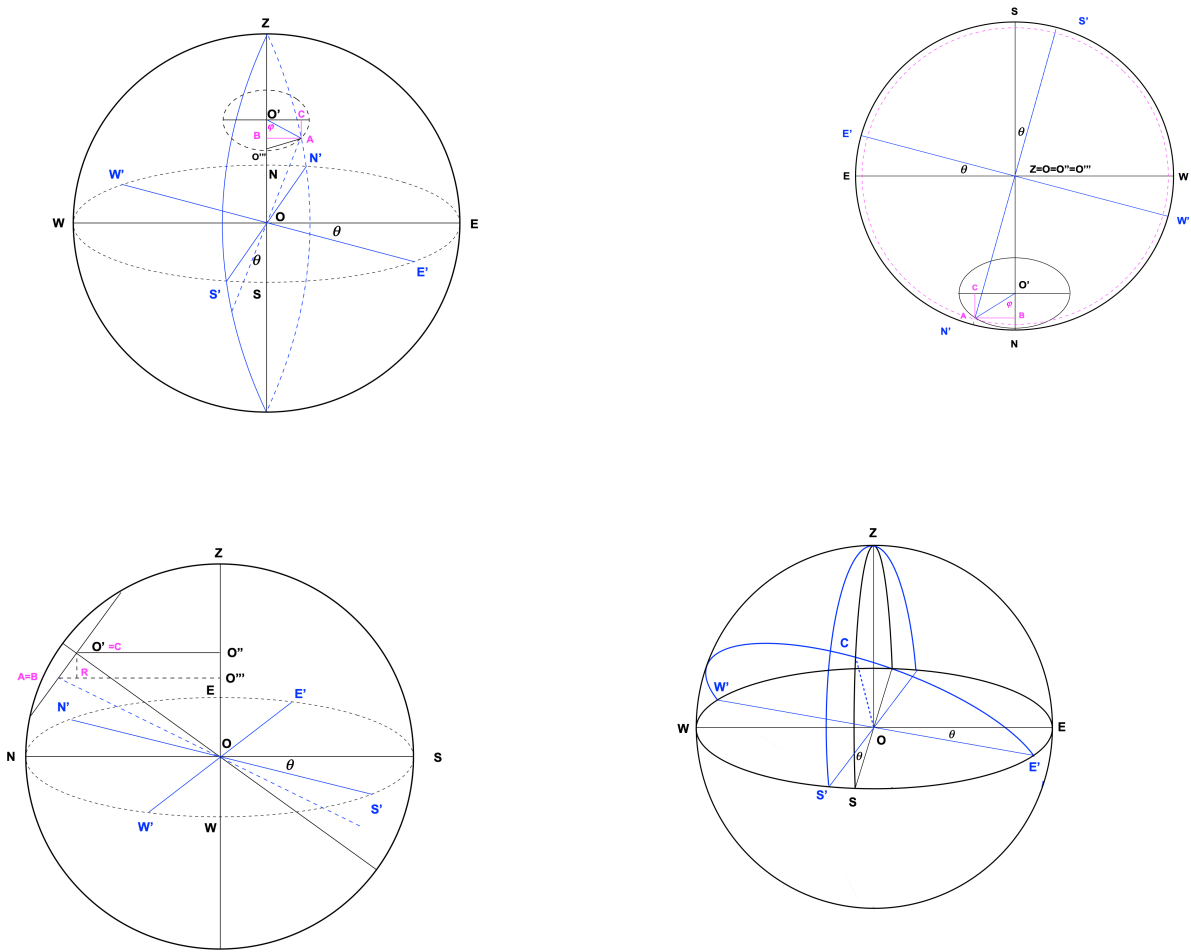

***Figure 3***. *Image of the celestial sphere as seen from the local south direction S (Top left), from the local zenith Z (Top right) and from the local west direction W (Bottom left). Bottom right shows the celestial sphere and the local south meridian (black SZ curve) along with ecliptic half circle above the local horizon (E'CW' blue curve) and the ecliptic meridian perpendicular to E'W' (S'Z blue curve).*

*The circle E-S-W-N is the local horizon. Blue axis E'W' indicates the intersection between the ecliptic plane and the horizontal plane; S' and N' are the directions where the ecliptic is highest (S') and lowest (N') relative to the horizon. The circle centered on O' indicates the path of the ecliptic N pole (located in A) in its rotation around the terrestrial axis (OO'). O" is the projection of O' on the Earth's axis, while O''' is the projection of point A on the terrestrial axis. The angle θ is the angle between the local south direction and the meridian of the highest point of the ecliptic, and φ is the Right Ascension coordinate shifted by 6h (or 90 degrees).*